\documentclass[11pt,a4paper]{article}
\usepackage{amsmath}
\usepackage{amssymb}
\usepackage{amsthm}
\usepackage{times}
\usepackage{graphicx}
\newcommand{\integers}{\mathbb{Z}}
\newcommand{\rationals}{\mathbb{Q}}
\newcommand{\reals}{\mathbb{R}}

\usepackage[hang,stable]{footmisc}
\begin{document}
\ifx\href\undefined\else\hypersetup{linktocpage=true}\fi
% Body
\title{Max Born's\\ \emph{Vorlesungen \"uber Atommechanik, Erster Band}}
\author{Domenico Giulini\\
Institute for Theoretical Physics, University of Hannover\\ 
Appelstrasse\,2, D-30167 Hannover, Germany\\ and\\
Center of Applied Space Technology and Microgravity (ZARM),\\
University of Bremen, Am Fallturm\,1, D-28359 Bremen, Germany.\\
\texttt{giulini@itp.uni-hannover.de}}

\date{}

\maketitle

\begin{abstract}
\noindent
A little more than half a year before Matrix Mechanics was born,
Max Born finished his book \emph{Vorlesungen \"uber Atommechanik, 
Erster Band}, which is a state-of-the-art presentation of 
Bohr-Sommerfeld quantisation.%
\footnote{As usual, I use the term ``Bohr-Sommerfeld quantisation'' 
throughout as shorthand for what probably should be called 
Bohr-Ishiwara-Wilson-Planck-Sommerfeld-Epstein-Schwarzschild$\cdots$ 
quantisation.} This book is remarkable for its epistemological as 
well as technical aspects. In this contribution I highlight one 
aspect in each of these two categories, the first being concerned 
with the r\^ole of axiomatisation in the heuristics of physics, 
the second with the problem of quantisation proper before Heisenberg
and Schr\"odinger.  

\setcounter{tocdepth}{2}
\setcounter{secnumdepth}{2}

{\footnotesize \tableofcontents}
\newpage 

\end{abstract}

\section{Outline}
\label{sec:Introduction}
Max Born's monograph \emph{Vorlesungen \"uber Atommechanik, Erster Band,}
was published in 1925 by Springer Verlag (Berlin) as volume II in the 
Series \emph{Struktur der Materie}~\cite{Born:VorlesungenAtommechanik}. 
The second volume appeared in 1930 as \emph{Elementare Quantenmechanik}, 
coauthored by Pascual Jordan, as volume~IX in the same series. 
Here the authors attempt to give a comprehensive and 
self-contained account of Matrix 
Mechanics~\cite{Born.Jordan:ElementareQuantenmechanik}.
The word ``elementare'' in the title alludes, in a sense, to the 
logical hierarchy of mathematical structures and is intended 
to mean ``by algebraic methods (however sophisticated) 
only'', as opposed to Schr\"odinger's wave mechanics, which uses 
(non elementary) concepts from calculus. Since by the end of 1929 
(the preface is dated December 6th 1929) several comprehensive 
accounts of wave mechanics had already been 
published\footnote{Born and Jordan mention the books by 
Haas~\cite{Haas:MateriewellenUndQM}, Sommerfeld~\cite{Sommerfeld:AUS-II},
de\,Broglie~\cite{Broglie:Wellenmechanik}, and
Frenkel\cite{Frenkel:EinfuehrungWellenmechanik}.},
the authors felt that it was time to do the same for Matrix
Mechanics. 

Here I will focus entirely on the first volume, which
gives a state-of-the-art account of Bohr-Sommerfeld quantisation
from the analytic perspective. One might therefore suspect 
that the book had almost no impact on the post-1924 
development\footnote{The preface is dated November 1924.} of 
Quantum Mechanics proper, whose 1925-26 breakthrough did not 
originate from yet further analytical refinements of 
Bohr-Sommerfeld theory.\footnote{A partial revival and 
refinement of Bohr-Sommerfeld quantisation set in during the 
late 1950s, as a tool to construct approximate solutions to 
Schr\"odinger's equation, even for non-separable 
systems~\cite{Keller:1958}; see 
also~\cite{Gutzwiller:ChaosClassQuantPhys}. Ever since it 
remained an active field of research in atomic and molecular 
physics.} But this would 
be a fruitless approach to Born's book, which is truly 
remarkable in at least two aspects: First, for its presentation 
of analytical mechanics, in particular Hamilton-Jacobi theory 
and its applications to integrable systems as well as 
perturbation theory and, second, for its epistemological 
orientation; and even though it is very tempting indeed to 
present some of the analytic delicacies that Born's book has 
to offer, I feel equally tempted to highlight some of the 
epistemological aspects, since the latter do not seem to we widely 
appreciated. In  contrast, Born's book is often cited and 
praised in connection with Hamilton and Hamilton-Jacobi theory, 
like e.g. in the older editions of Goldstein's book on classical
mechanics.\footnote{In the latest editions (2002 English, 
2006 German) the author's seem to have erased all references 
to Born's book.} 

\section{Structure of the Book}
\label{sec:Structure}
The book is based on lectures Born had given in the winter semester 
1923/24 at the University of G\"ottingen and written with the help 
of Born's assistant Friedrich Hund, who wrote substantial parts 
and contributed important mathematical results (uniqueness of 
action-angel variables). Werner Heisenberg outlined some paragraphs, 
in particular the final ones dealing with the Helium atom.
The text is divided into 49 Sections, grouped into 5 chapters,
and a mathematical appendix, which together amount to almost 
350 pages. It may be naturally compared and contrasted with 
Sommerfeld's \emph{Atombau and Spektrallinien\,I} , which has 
about twice the number of pages. As already said, Born's text 
is today largely cited and remembered (if at all!) for its 
presentation of Hamilton-Jacobi theory and perturbation theory 
(as originally developed for astronomical problems), which 
is considered comprehensive and most concise, though today 
one would approach some of the material by more geometric 
methods (compare Arnold's book \cite{Arnold:Mechanics} or 
that of Abraham \& Marsden~\cite{Abraham.Marsden:Mechanics}).

The list of contents on the level of chapters is as follows:
\begin{itemize} 
\item[Intro.:]
Physical Foundations (3 sections, 13 Pages)
\\[-4.0ex]
\item[Ch.1:]
Hamilton-Jacobi Theory (5 sections, 23 pages)
\\[-4.0ex]
\item[Ch.2:]
Periodic and multiple periodic motions (12 sections, 81 pages)
\\[-4.0ex]
\item[Ch.3:]
Systems with a single valence (`light') electron 
(19 sections, 129 pages)
\\[-4.0ex]
\item[Ch.4:]
Perturbation theory (10 sections, 53 pages)
\end{itemize}

Both \emph{Vorlesungen \"uber Atommechanik} were reviewed  by 
Wolfgang Pauli for \emph{Die Naturwissenschaften}. In his Review 
of the first volume, young Pauli emphasised in a somewhat pointed 
fashion its strategy to apply mechanical principles to special 
problems in atomic physics, of which he mentioned the following
as essential ones: Keplerian motion and the influence it receives 
from relativistic mass variations and external fields, general 
central motion (Rydberg-Ritz formula), diving orbits 
[``Tauchbahnen''], true principal quantum numbers of optical 
terms, construction of the periodic system according to Bohr, 
and nuclear vibrations and rotation of two-atomic molecules. 
He finally stresses the elaborateness of the last chapter on 
perturbation theory,
\begin{quote}
``...of which one cannot say, that the invested effort corresponds 
to the results achieved, which are, above all, mainly negative 
(invalidity of mechanics for the Helium atom). Whether this method
can be the foundation of the true quantum theory of couplings, as 
the author believes, has to be shown by future developments. May this 
work itself accelerate the development of a simpler and more unified 
theory of atoms with more than one electron, the manifestly unclear 
character as of today is clearly pictured in this chapter.''  
(\cite{Pauli_Rev:1925}, p.\,488)
\end{quote}

As an amusing aside, this may  be compared with Pauli's review 
of the second volume, which showed already considerably more of 
his infamous biting irony.  Alluding to Born's as well as Born's \& 
Jordan's own words in the introductions to volume 1 and 2 
respectively, Pauli's review starts 
with: 
\begin{quote}
``This book is the second volume of a series, in which each time the 
aim and sense [Ziel und Sinn] of the nth volume is made clear by 
the virtual existence of the (n+1)st.''
(\cite{Pauli_Rev:1930}, p.\,602)
\end{quote}
 Having given no recommendation, the review then ends with:
\begin{quote} 
``The making [Ausstattung] of the book with respect to print and 
paper is excellent [vortrefflich]''.
(\cite{Pauli_Rev:1930}, p.\,602)

\end{quote}

\newpage
%\begin{figure}[h]
\begin{center}
\includegraphics[width=0.9\linewidth]{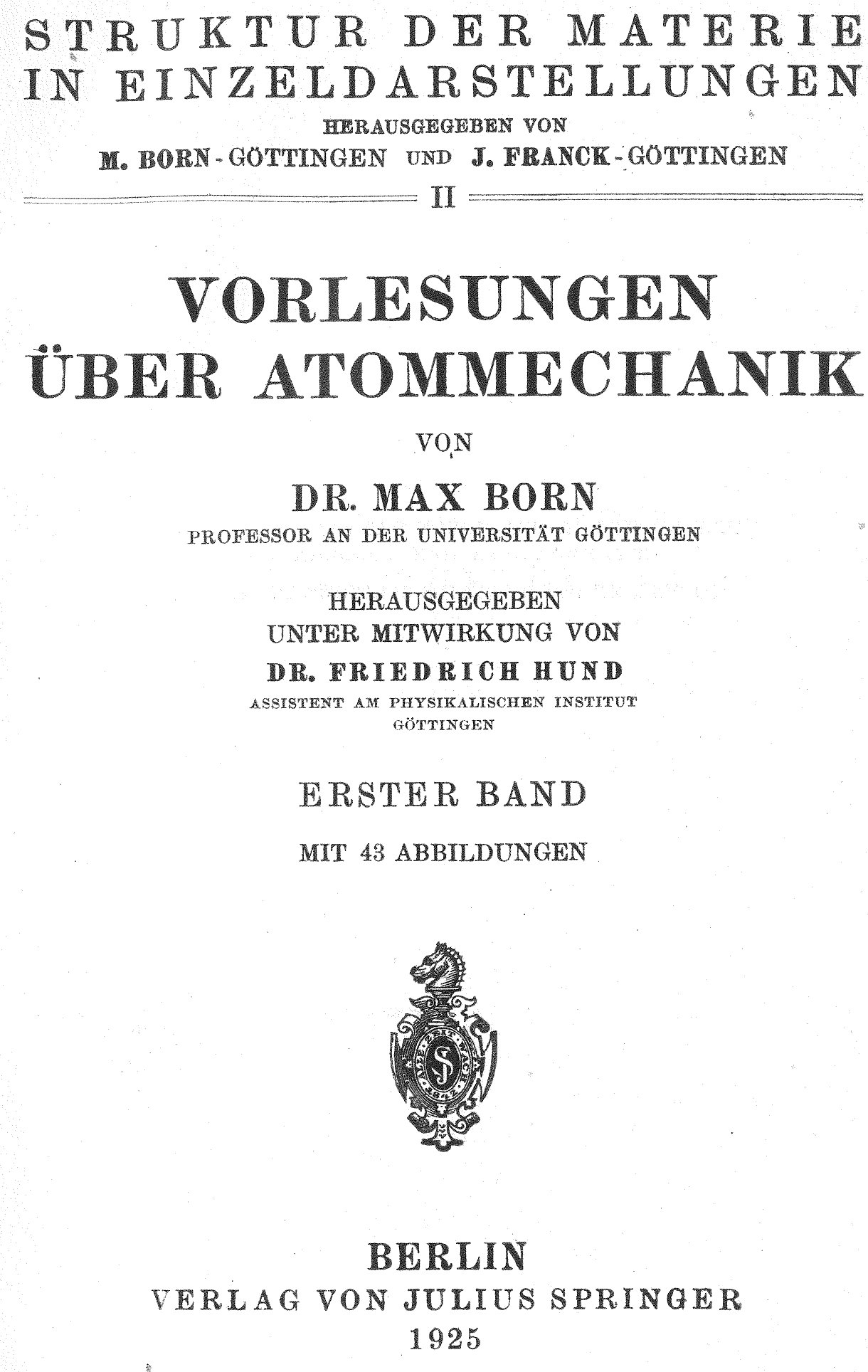}
\end{center}

\smallskip
\begin{center}
{\Large Fig.\,1: Title Page}
\end{center}
%\caption{\label{fig:BornCover} Cover page} 
%\end{figure}
\newpage
%\begin{figure}[h]
\begin{center}
\includegraphics[width=1.0\linewidth]{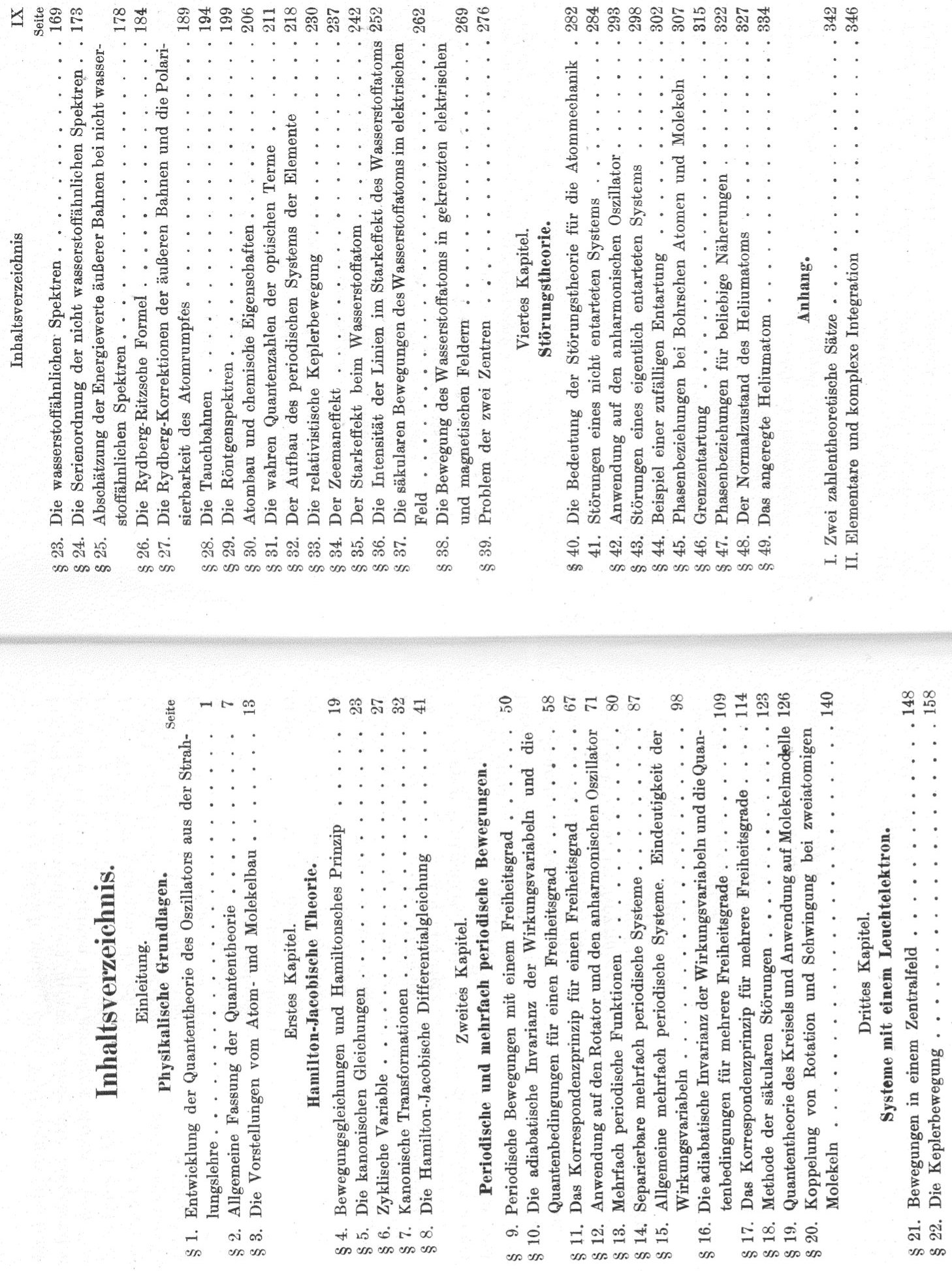}
\end{center}

\smallskip
\begin{center}
{\Large Fig.\,2: Table of Contents}
\end{center}
%\caption{\label{fig:BornTOC} Table of contents} 
%\end{figure}
\newpage

\section{Born's pedagogy and the heuristic r\^ole 
of the deductive/axiomatic method}
\label{sec:BornsPedagogy}
\subsection{Sommerfeld versus Born}
\label{sec:Sommerfeld versus Born}
Wilhelm von Humboldt's early 19th-century programmatic vision 
of an intimate coexistence and cross fertilisation of teaching 
and research soon became a widely followed paradigm for 
universities in Prussia, other parts of Germany, and around 
the World. And even though it is clear from experience that 
there cannot be a general rule saying that the best researchers 
make the best teachers and vice versa, Humboldt's programme has 
nevertheless proven extremely successful. In fact, outstanding 
examples for how to suit the action to the word are provided by 
the Munich and G\"ottingen schools of Quantum Physics during 
the post-World-War-I period. Their common commitment to the 
``Humboldian Ideal'', with action speaking louder than words, 
resulted in generations of researchers and teachers of highest 
originality and quality. What makes this even more convincing 
is the impression that this was not achieved on account of 
personal individuality; quite the contrary. Sommerfeld in 
Munich, for example, is well known to have had an extraordinary 
fine sense for the gifts of each individual students and how to 
exploit it in an atmosphere of common scientific 
endeavour~\cite{Seth:CraftingTheQuantum}. 
Similar things can be said of Max Born in G\"ottingen, though perhaps 
not quite as emphatic. Born's style was slightly less adapted 
to the non-systematic approaches of scientific greenhorns, whereas
Sommerfeld would appreciate any new ideas and tricks, if only 
for the purpose of problem solving. For Sommerfeld, teaching the 
art of problem solving was perhaps the single most important 
concern in classes and seminars~\cite{Seth:CraftingTheQuantum}. 
Overly tight and systematic expositions are not suited for that 
purpose. This point was often emphasised by Sommerfeld, for example 
right at the beginning of his classic five-volume ``Lectures on 
Theoretical Physics''. The first volume is called ``Mechanics'', not 
``Analytical Mechanics'' as Sommerfeld stresses in a one-page 
preliminary note that follows the preface, since 
\begin{quote}
``This name [analytical mechanics] originated in the grand work 
of Lagrange's of 1788, who wanted to cloth all of mechanics in
a uniform language of formulae and who was proud that one would 
not find a single figure throughout his work. We,  in contrast,
will resort to intuition [Anschauung] whenever possible and 
consider not only astronomical but also physical and, to a 
certain extent, technical applications.'' 
\end{quote}
The preface already contains the following programmatic 
paragraph, which clearly characterises Sommerfeld's approach 
to teaching in general: 
\begin{quote}
``Accordingly, in print [as in his classes; D.G.] I will not 
detain myself with the mathematical foundations, but proceed 
as rapidly as possible to the physical problems themselves. 
I wish to supply the reader with a vivid picture of the highly 
structured material that comes within the scope of theory from 
a suitable chosen mathematical and physical vantage point. 
May there, after all, remain some gaps in the systematic 
justification and axiomatic consistency. In any case during my 
lectures I did not want to put off my students with tedious 
investigations of mathematical or logical nature and distract
them from the physically interesting. This approach has,
I believe, proven useful in class and has been maintained in 
the printed version. As compared to the lectures by Planck, 
which are impeccable in their systematic structure, I believe 
I can claim a greater variety in the material and a more 
flexible handling of the mathematics.''        
\end{quote}
This pragmatic paradigm has been taken over and perfected by 
generations of theoretical physicists; just think of the 
10-volume lecture courses by Landau and Lifshitz, which is 
still in print in many languages and widely used all over 
the world. 

There are many things to be said in favour of this pragmatic 
approach. For one thing, it takes account of the fact that 
understanding is a cyclic process. Every student knows that 
one has to go over the same material again and again in order 
to appreciate the details of the statements, its hidden 
assumptions, and the intended range of validity. Often on 
one's $n$th iteration one discovers new aspects, in view 
of which one's past understanding is revealed as merely 
apparent and ill based. Given that we can almost never be 
sure for this not to happen again, one might even be tempted 
to measure one's own \emph{relative} degree of understanding 
by the number of times this has already happened in the past. 
From that perspective, the pragmatic approach seems clearly 
much better suited, since it does not pretend the fiction 
of an ultimate understanding. Being able to solve concrete 
problems sounds then like a reasonable and incorruptible 
criterion. 

However, as Thomas Kuhn pointed out long ago, well 
characterised (concrete) problems, also called ``puzzles''
by him, must be supplied by paradigms to which the working scientists 
adhere. If concrete problems become critically severe, 
with eventually all hopes for solutions under the current 
paradigm fading away, further puzzle-solving activities 
will sooner or later decouple from further progress. 
The crucial question then is: Where can seeds for further 
progress be found and how should they be planted? 

It is with regard to  this question that I see a clear 
distinction between the approaches of Born and Sommerfeld. 
Sommerfeld once quite frankly admitted to Einstein:
\begin{quote}
``Everything works out all right [klappt] and yet remains 
fundamentally unclear. I can only cultivate [f\"ordern] 
the techniques of the quanta, you have to provide 
your philosophy.''
(\cite{Hermann:EinsteinSommerfeldBriefwechsel}, p.\,97).
\end{quote}
The planting of seeds could start with simple axioms 
in a well defined mathematical framework. But even that 
might turn out to be premature. Heisenberg is one of the 
figures who repeatedly expressed the optimistic view that
physical problems can be ``essentially'' solved while 
being still detached from such a framework. In connection 
with his later search for a unified field theory of 
elementary particles he said in the preface to his 
textbook on that matter: 
\begin{quote}
``At the current status of the theory it would be premature 
to start with a system of well defined axioms and then 
deduce from them the theory by means of exact mathematical  
methods. What one needs is a mathematical description which 
adequately describes the experimental situation, which 
does not seem to contain contradictions and which, therefore,
might later be completed to an exact mathematical scheme. 
History of physics teaches us that, in general, a new theory 
can be phrased in a precise mathematical language only after 
all essential physical problems have been solved.''%
(\cite{Heisenberg:EinheitlicheFeldtheorie}, p.\,VI)
\end{quote}
It seems even more obvious that in phases of paradigmatic 
uncertainty not much help can be expected from attempts to 
establish an axiomatic framework for the doomed theory. 
And yet, quite surprisingly, this is precisely what Born did,
as we shall see in the next subsection.
 
In a letter to Paul Ehrenfest from 1925, Einstein divided 
the community of physicists into the \emph{Prinzipienfuchser}
and the \emph{Virtuosi} (\cite{Seth:CraftingTheQuantum}, p.\,186).%
\footnote{As Seth already remarked in Note 29 to Chapter 6 
of~\cite{Seth:CraftingTheQuantum}, \emph{Prinzipienfuchser} is 
nearly untranslatable. Existing compound words are Pfennigfuchser 
(penny pincher) and Federfuchser (pedant) (not `Pfederfuchser',
as stated in~\cite{Seth:CraftingTheQuantum}, which does not 
exist).}
Einstein saw Ehrenfest, Bohr, and himself in the first category and 
named Debey and Born as members of the latter one. Virtuosity here 
refers to the high mathematical and calculational abilities, any 
encounter of which results in mental depression on the side of the 
Prinzipienfuchser, as Einstein concedes to Ehrenfest who first 
complained about this effect. However, Einstein adds that opposite 
effect exists, too.  

This dichotomy is certainly not strictly exclusive. An obvious 
example of somebody who could with equal right be located in both 
camps is Wolfgang Pauli. But also Born lives in both camps and can 
be best described, I think, as a Prinzipienfuchser amongst the 
Virtuosi. The principles about which he is so much concerned arise 
within the attempt to find a logical basis from which the physically
relevant can be deduced without ambiguity, rather than just apply 
clever tricks. This difference to the Sommerfeld school has once 
been expressed by Heisenberg in an interview to Thomas Kuhn from 
February 15th 1963: 
\begin{quote}
``In Sommerfeld's institute one learned to solve special problems; 
one learned the tricks, you know. Born took it much more 
fundamentally, from a very general axiomatic point of view. So only 
in G\"ottingen did I really learn the techniques well. Also in this 
way Born's seminar was very helpful for me. I think from this Born 
seminar on I was able really to do perturbation calculations with 
all the rigour which was necessary to solve such problems.''
(Quoted in \cite{Seth:CraftingTheQuantum}, p.\,58).
\end{quote}       
Let us now see how Born himself expresses the heuristic value 
of the axiomatic method in times of uncertainty.

\subsection{A remarkable introduction}
\label{sec:RemarkableIntroduction}
One third through the book, Born recalls the basic idea 
of `Quantum Mechanics' in the following way (the emphases
are his; the German original of various terms and phrases 
are included in square brackets): 
\begin{quote}
``Once again, we summarise the basic idea of Quantum Mechanics, 
as developed so far: \emph{For a given Model [Modell] we calculate 
the totality of all  motions} (which are assumed to be multiply 
periodic) \emph{according to the laws of Classical Mechanics}
(neglecting radiation damping); \emph{the quantum conditions select 
a discrete subset from this continuum of motions.} The \emph{energies} 
of the selected motions shall be the \emph{true} [wirkliche] 
ones, as measurable by electron collision, and the energy 
differences shall, according to Bohr's frequency condition, 
correspond [zusammenh\"angen] with the \emph{true} [wirklichen] 
\emph{light frequencies}, as observed in the spectrum. Besides 
frequencies, the emitted light possesses the observable 
properties of intensity, phase, and state of polarisation, 
which are only approximately accounted for by the theory 
(\S\,17). These exhaust the observable properties of the motion 
of the atomic system. However, our computation assigns 
\emph{additional properties} to it, namely orbital frequencies 
and distances, that is, the course [Ablauf] of motion in time. 
It seems that these quantities are, as a matter of principle, 
not accessible to observation.\footnote{Here Born adds the 
following footnote: ``Measurements of atomic radii and the 
like do not lead to better approximations to reality 
[Wirklichkeit] as, say, the coincidence between orbital and 
light frequencies.''} Therewith we arrive at the following 
judgement [Urteil], that \emph{for the time being our procedure 
is just a formal computational scheme} which, for certain cases,
allows us to replace the still unknown quantum laws by 
computations on a classical basis [auf klassischer Grundlage].
Of these true [wahren] laws we would have to require, that 
they only contain relations between observable quantities, 
that is, energy, light frequencies, intensities, and phases. 
As long as these laws are still unknown, we have to always 
face the possibility that our provisional quantum rules 
will fail; one of our main tasks will be to delimit [Abgrenzen]
the validity of these rules by comparison with experience.''
(\cite{Born:VorlesungenAtommechanik}, p.\,113-114)
\end{quote}
As an (obvious) side remark, we draw attention to the 
similarity between Born's formulations in the second half 
of the above cited passage and Heisenberg's opening 
sentences of his Umdeutung paper~\cite{Heisenberg:1925}.

Born's book attempts an axiomatic-deductive approach to 
Bohr-Sommerfeld quantisation. This might seem  totally 
misguided at first, as one should naively think that such 
a presentation only makes sense \emph{after} all the essential 
physical notions and corresponding mathematical structures have 
been identified. Certainly none of the serious researchers at 
the time believed that to be the case for Bohr-Sommerfeld
quantisation, with Born making no exception as we have just 
seen from his outline and judgement cited above. So what is 
Born's own justification for such an attempt? This he 
explains in his introduction to the book, where he takes a 
truly remarkable heuristic attitude. I found it quite 
inappropriate to interfere with his words, so I will now 
largely quote from that introduction [the translation is mine]: 
\begin{quote}
``The title `Atommechanik' of this lecture, which I delivered
in the winter-semester 1923/24 in G\"ottingen, is formed after 
the label `Celestial Mechanics'. In the same way as the latter 
labels that part of theoretical astronomy which is concerned 
with the calculation of trajectories of heavenly bodies 
according to the laws of mechanics, the word `Atommechanik' 
is meant to express that here we deal with the facts of atomic 
physics from the particular point of view of applying mechanical 
principles. This means that we are attempting a deductive 
presentation of atomic theory. The reservations, that the theory is 
not sufficiently mature [reif], I wish to disperse with 
the remark that we are dealing with a test case [Versuch], 
a logical experiment, 
the meaning of which just lies in the determination of the limits to
which the principles of atomic- and quantum physics succeed, and to 
pave the ways which shall lead us beyond that limits. I called this
book `Volume\,I' in order to express this programme already in the 
title; the second volume shall then contain a higher approximation 
to the `final' mechanics of atoms. 

I am well aware that the promise of such a second volume is 
daring [k\"uhn]; since 
presently we have only a few hints as to the nature of the deviations 
that need to be imposed onto the classical laws in order to explain 
the atomic properties. To these hints I count first  of all Heisenberg's 
rendering of the laws of multiplets and anomalous Zeeman effect, 
the new radiation theory of Bohr, Kramers, and Slater, the ensuing 
Ans\"atze of Kramers for a quantum-theoretic explanation of the 
phenomena of dispersion, and also some general considerations 
concerning the adaptation of perturbation theory to the quantum 
principles, which I recently communicated. But all this material, 
however extensive it might be, does not nearly suffice to shape a 
deductive theory from it. Therefore, the planned `2.~Volume' might 
remain unwritten for many years to come; its virtual existence may, 
for the time being, clarify the aim and sense [Ziel und Sinn]
of this book.[...]'' 
(\cite{Born:VorlesungenAtommechanik}, p.\,V-VI)
\end{quote}

 Born continues and explicitly refers (and suggests the 
reading of) Sommerfeld's Atombau und Spektrallinien, 
almost as a prerequisite for a successful study of his 
own book. But he also stresses the difference which, in part, 
lies in the deductive approach:

\begin{quote}
``For us the mechanical deductive approach always 
comes first [steht \"uberall obenan]. Details of 
empirical facts will only be given when they 
are essential for the clarification, the support, or 
the refutation of theoretical strings of thought
[Gedankenreihen].'' 
(\cite{Born:VorlesungenAtommechanik}, p.\,VI)
\end{quote}

But, Born continues, there is a second difference to Atombau 
und Spektrallinien, namely with respect to the foundations 
of Quantum Theory, where 

\begin{quote}
...``differences in the emphasis of certain features 
[Z\"uge] are present; but I leave it to the author to 
find these out by direct comparison. As regards the 
relation of my understanding to that of Bohr and his 
school, I am not not aware of any significant opposition. 
I feel particularly sympathetic with the Copenhagen researchers 
in my conviction, that it is a rather long way to go to a 
`final quantum theory'.''
(\cite{Born:VorlesungenAtommechanik}, p.\,VI)
\end{quote}

It would be an interesting project to try to work out the 
details of the `second difference', concerning the foundations 
of Quantum Theory, by close comparison of Born's text with 
Atombau und Spektrallinien. Later, as we know, Born conceptually 
favoured the more abstract algebraic approach (Heisenberg) 
against the more `anschauliche' wave-theoretic picture, quite 
in contrast to Sommerfeld, who took a more pragmatic stance.
Born's feeling that this conceptual value should receive a 
stronger promotion, for it is blurred by the semi-anschauliche 
picture of waves travelling in (high dimensional) configuration 
space, is clearly reflected in the second volume, as well as in 
later publications, like in the booklet by him and Herbert 
Green of 1968 on ``matrix methods in quantum mechanics''. This 
split attitude is still very much alive today, though it is 
clear that in terms of calculational economy wave mechanics 
is usually preferred. 

Born ends his introduction by acknowledging the help of several 
people, foremost his assistant Friedrich Hund for his 
``devoted collaboration'':
\begin{quote}
Here I specifically mention the theorem concerning the 
uniqueness of action-angle variables which, 
according to my view, lies at the foundation of 
\emph{today's} quantum theory; the proof worked out by 
Hund forms the centre [Mittelpunkt] of the second 
chapter (\S\,15).''
(\cite{Born:VorlesungenAtommechanik}, p.\,VII)
\end{quote}
Hund is also thanked for the presentation of Bohr's theory of 
periodic systems. Heisenberg is thanked for his advice and for 
outlining particular chapters, like the last one on the Helium atom. 
L.\,Nordheim's help with the presentation of perturbation theory is 
acknowledged and H.\,Kornfeld for checking some calculations. Finally 
F.\,Reiche H.\,Kornfeld and F.\,Zeilinger are thanked for helping 
with corrections.

\section{On technical issues: What is quantisation?}
\label{sec:QuantisationRules}
A central concern of Born's book is the issue of quantisation 
rules, that is: How can one \emph{unambiguously} generalise 
\begin{equation}
\label{eq:QuantRuleOneDegFreed}
J:=\oint p\,dq=nh
\end{equation}
to systems with more than one degrees of freedom?
The history of attempts to answer this question is 
interesting but also rather intricate, and involves 
various suggestions by 
Ishiwara~\cite{Ishiwara:1915}, 
Wilson~\cite{Wilson:1915}, 
Planck~\cite{Planck:1916},
Sommerfeld~\cite{Sommerfeld:1916b},  
Schwarzschild~\cite{Schwarzschild:1916c}, 
Epstein \cite{Epstein:1916a,Epstein:1916c}, 
and, last not least, the somewhat singular paper by 
Einstein from 1917 on 
``The Quantum Theorem of Sommerfeld and Epstein'' 
(\cite{Einstein:CP}, Vol.\,6, Doc.\,45, pp.\,556-567),
to which we turn below. These papers have various 
logical dependencies and also partially differ in subtle 
ways. Leaving aside Einstein's paper for the moment, the 
rule that emerged from the discussions looked innocently 
similar to (\ref{eq:QuantRuleOneDegFreed}), namely 
\begin{equation}
\label{eq:QuantRuleManyDegFreed}
J_k:=\oint p_k\,dq_k=n_kh\qquad\text{(no summation over k)}
\end{equation}
where $k=1,2,\cdots,s$ labels the degrees of freedom 
to be quantised, which need not necessarily exhaust all 
physical degrees of freedom, of which there are $f\geq s$,
as we shall discuss below.\footnote{In (\ref{eq:QuantRuleManyDegFreed})
as well as in all formulae to follow, we shall never make use of 
the summation convention.} Here we adopt the notation from 
Born's book, where $(q_1,\cdots,q_f;p_1,\cdots, p_f)$ are 
the generalised coordinates (configuration variables) and 
momenta respectively. The apparent simplicity of 
(\ref{eq:QuantRuleManyDegFreed}) is deceptive though. One thing 
that needs to be clarified is the domain of integration, here 
implicit in the $\oint$-symbol. It indicates that the 
integration over $q_k$ is to be performed over a full 
periodicity interval of \emph{that} configuration variable.
In Sommerfeld's words (his emphases):
\begin{quote}
``\emph{Each coordinate shall be extended over the full range necessary 
to faithfully label the phase of the system.} For a cyclic azimuth 
in a plane this range is $0$ to $2\pi$, for the inclination in space 
(geographic latitude $\theta$) twice the range between $\theta_{\mathrm{min}}$  
and $\theta_{\mathrm{max}}$, for a radial segment $r$ [Fahrstrahl] 
likewise twice the covered interval from $r_{\mathrm{min}}$  
to $r_{\mathrm{max}}$ for the motion in question.''
(\cite{Sommerfeld:1916b}, p.\,7)
\end{quote} 
Another source of uncertainty concerns the choice of canonical 
coordinates in which (\ref{eq:QuantRuleManyDegFreed}) is meant to 
hold. Again in Sommerfeld's words of his comprehensive 
1916 account:
\begin{quote}
``Unfortunately a general rule for the choice of coordinates
can hardly be given; it will be necessary to collect further 
experience by means of specific examples. In our problems 
it will do to use (planar and spatial) polar coordinates.
We will come back to a promising rule of Schwarzschild and 
Epstein for the choice of coordinates in \S\,10.''
(\cite{Sommerfeld:1916b}, p.\,6)
\end{quote}
The rule that Epstein and independently Schwarzschild formulated 
in their papers dealing with the Stark effect (\cite{Epstein:1916a} 
and \cite{Schwarzschild:1916c} respectively, compared by Epstein 
in \cite{Epstein:1916c} shortly after Schwarzschild's death) 
is based on the assumptions that, first, Hamilton's equations 
of motion 
\begin{equation}
\label{eq:HamiltonEqOfMotion}
{\dot q}_k=\partial H/\partial p_k\,,\qquad 
{\dot p}_k=-\partial H/\partial q_k\,,
\end{equation}
for time independent Hamiltonians 
$H(q_1,\cdots,q_f;p_1,\cdots,p_f)$ are solved by means 
of a general solution 
$S(q_1,\cdots,q_f;\alpha_1,\cdots ,\alpha_f)$ for the 
Hamilton-Jacobi equation
\begin{equation}
\label{eq:HamiltonJacobiEq}
H\left(q_1,\cdots,q_f;
\frac{\partial S}{\partial q_1},\cdots,
\frac{\partial S}{\partial q_f}\right)=E\,,
\end{equation}
where $p_k=\partial S/\partial q_k$ and $\alpha_1,\cdots,\alpha_f$
are constants of integration on which the energy $E$ depends. 
Second, and most importantly, that this solution is obtained 
by separation of variables:
\begin{equation}
\label{eq:SepVar}
S(q_1,\cdots,q_f;\alpha_1,\cdots ,\alpha_f)
=\sum_{i=1}^f S_i(q_i;\alpha_1,\cdots,\alpha_f)\,.
\end{equation}
Note that this in particular implies that 
$p_k=p_k(q_k;\alpha_1,\cdots,\alpha_f)$, i.e. the 
$k$th momentum only depends on the $k$th configuration 
variable and the $f$ constants of integration 
$\alpha_1,\cdots,\alpha_f$. This is indeed necessary for 
(\ref{eq:QuantRuleManyDegFreed}) to make sense, since the right 
hand side is a constant and can therefore not be be meaningfully 
equated to a quantity that depends non trivially on phase space. 
Rather, the meaning of (\ref{eq:QuantRuleManyDegFreed}) is to 
select a subset of solutions through equations for the $\alpha$s. 
However, separability is a very strong requirement indeed
which, in particular, requires the integrability of the 
dynamical system in question, a fact to which only Einstein 
drew special attention to in his paper~(\cite{Einstein:CP}, 
Vol.\,6, Doc.\,45, pp.\,556-567), as we will discuss in more 
detail below. In fact, integrability is manifest once the 
$J_1,\cdots,J_f$ have been introduced as so-called 
``action variables'', which are conjugate to some 
``angle variables'' $w_1,\cdots,w_f$; for then the action 
variables constitute the $f$ observables in involution, i.e. 
their mutual Poisson brackets obviously all vanish.%
\footnote{The implication of integrability for separability 
is far less clear; compare, e.g., 
\cite{Gutzwiller:ChaosClassQuantPhys}. Classic results 
concerning sufficient conditions for separability were 
obtained by St\"ackel; see \cite{Charlier:Himmelsmechanik-1})}  

But even if we swallow integrability as a \emph{conditio sine qua non}, 
does separability ensure uniqueness? What is the strongest uniqueness 
result one can hope for? Well, for (\ref{eq:QuantRuleManyDegFreed}) 
to make sense, any two allowed (by conditions yet to be formulated) 
sets of canonical coordinates $(q_i,p_i)_{i=1\cdots n}$ and 
$(\bar q_i,\bar p_i)_{i=1\cdots n}$ must be such that the $(J_k/h)$s 
(calculated according to (\ref{eq:QuantRuleManyDegFreed})) 
are integers if and only if the $(\bar J_k/h)$s are. This is 
clearly the case if the allowed transformations are such 
that among the action variables $J_k$ they amount to linear 
transformations by invertible integer-valued 
matrices:\footnote{Note that the inverse matrices must also be 
integer valued; hence the matrices must have determinant equal 
to $\pm1$.}
\begin{subequations}
\label{eq:IntegerTransformationVar}
\begin{equation}
\label{eq:IntegerTransformationActionVar}
\bar J_k=\sum_{l=1}^f\tau_{lk}J_l\quad(\tau_{lk})\in\mathrm{GL}(f,\integers)\,.
\end{equation} 
Here $\mathrm{GL}(f,\integers)$ is the (modern) symbol for the 
group of invertible $f\times f$ -- matrices with integer
entries. The most general transformations for the angle 
variables compatible with (\ref{eq:IntegerTransformationActionVar})
are 
\begin{equation}
\label{eq:IntegerTransformationAngleVar}
\bar w_k=\sum_{l=1}^f\tau^{-1}_{kl}w_l+\lambda_k(J_1,\cdots,J_f)\,,
\end{equation} 
\end{subequations}
where the $\lambda_k$ are general (smooth) 
functions.\footnote{Our equation 
(\ref{eq:IntegerTransformationAngleVar}) differs in a 
harmless fashion from the corresponding equation\,(7) 
on p.\,102 of \cite{Born:VorlesungenAtommechanik}, 
which reads $w_k=\sum_{l=1}^f\tau_{kl}\bar w_l+\psi_k(J_1,\cdots,J_f)$,
into which our equation turns if we redefine the 
functions through $\psi_k=-\sum_{l=1}^f\tau_{kl}\lambda_l$.}

The task is now to carefully amend the Epstein-Schwarzschild 
condition of separability by further technical assumptions under 
which the transformations (\ref{eq:IntegerTransformationVar})
are the \emph{only} residual ones. The solution of this 
problem is presented in \S\,15 of Born's book, who 
acknowledges essential help with this by Friedrich Hund. 

Born also states that the technical conditions under which 
this result for multiply periodic systems can be derived 
were already given  in the unpublished thesis by 
J.M.\,Burgers~\cite{BurgersThesis:1918}, who is better 
known for his works on the adiabatic invariants. 
The arguments in Burger's thesis to show uniqueness are, 
according to Born, technically incomplete. 
The conditions themselves read as follows:
\begin{itemize}
\item[A]
The position of the system shall periodically depend on 
the angle variables $(w_1,\cdots,w_f)$ with primitive 
period 1.
\item[B]
The Hamiltonian is transformed in to a function $W$
depending only on the $(J_1,\cdots,J_f)$.%
\footnote{We follow Born's notation, according to which 
the Hamiltonian, considered as function of the action 
variables, is denoted by $W$.}
\item[C]
The phase-space function
\begin{equation}
\label{eq:BornCond-1}
S^*=S-\sum_{k=1}^f w_kJ_k\,,
\end{equation}
considered as function of the variables $(q,w)$, 
which generates the canonical transformation 
$(q,p)\mapsto(w,J)$ via
\begin{equation}
\label{eq:BornCond-2}
p_k= \frac{\partial S^*}{\partial q_k}\qquad
J_k=-\frac{\partial S^*}{\partial w_k}\,,
\end{equation}
shall also be a periodic function of the $w$s with period~1.
\end{itemize} 
A and B are immediately clear, but the more technical condition 
C is not. But, as Born remarks, A and B no not suffice to lead 
to the desired result. In fact, a simple canonical transformation 
$(w,J)\mapsto (\bar w,\bar J)$ compatible with A and B is 
\begin{equation}
\label{eq:BornCond-3}
\bar w_k=w_k+f_k(J_1,\cdots,J_f)\,,\quad
\bar J_k=J_k+c_k\,,
\end{equation}
where the $c_k$ are arbitrary constants. Their possible 
presence disturbs the quantisation condition, since 
$J_k$ and $\bar J_k$ cannot generally be simultaneously 
integer multiples of $h$. Condition C now eliminates this 
freedom. After some manipulations the following result 
is stated on p.\,104 of~\cite{Born:VorlesungenAtommechanik}:

\medskip
\noindent
\textbf{Theorem (Uniqueness for non-degenerate systems)}
\emph{
If for a mechanical system variables $(w,J)$ can be introduced 
satisfying conditions A-C, and if there exist no commensurabilities 
between the quantities
\begin{equation}
\label{eq:DefOfFrequencies}
\nu_k=\frac{\partial W}{\partial J_k}\,,
\end{equation} 
then the action variables $J_k$ are determined uniquely up 
to transformations of type (\ref{eq:IntegerTransformationActionVar})
[that is, linear transformations by $\mathrm{GL}(f,\integers)$]}.
\medskip

For the proof, as well as for the ensuing interpretation of 
the quantisation condition, the notions of \emph{degeneracy} 
and \emph{commensurability} are absolutely essential: An $f$-tuple
$(\nu_1,\cdots,\nu_f)$ of real numbers is called $r$-fold 
degenerate, where $0\leq r\leq f$, if there are $r$ but not 
$r+1$ independent integer relations among them, that is, if 
there is a set of $r$ mutually independent $f$-tuples 
$n^{(\alpha)}_1,\cdots ,n^{(\alpha)}_f$, $\alpha=1,\cdots,r$
of integers, so that $r$ relations of the form 
\begin{equation}
\label{eq:DefDegeneracy}
\sum_{k=1}^f n^{(\alpha)}_k\nu_k=0\,,\quad\forall \alpha=1,\cdots,r\,.
\end{equation}
hold, but there are no $r+1$ relations of this sort. 
The $f$-tuple is simply called degenerate if it is $r$-fold 
degenerate for some $r>0$. A relation of the form 
(\ref{eq:DefDegeneracy}) is called a commensurability. If no 
commensurabilities exist, the system called non-degenerate
or incommensurable.

It is clear that a relation of the form (\ref{eq:DefDegeneracy}) 
with $n_k^{(\alpha)}\in\integers$ exists if and only if it 
exists for $n_k^{(\alpha)}\in\rationals$ (rational numbers). 
Hence a more compact definition of $r$-fold degeneracy is 
the following: Consider the real numbers 
$\reals$ as vector space over the rational numbers $\rationals$
(which is infinite dimensional). The $f$ vectors $\nu_1,\cdots,\nu_f$
are $r$-fold degenerate if and only if their span is 
$s$-dimensional, where $s=f-r$.

Strictly speaking, we have to distinguish between \emph{proper}
[Born: ``eigentlich''] and \emph{improper} (or contingent) 
[Born: ``zuf\"allig''] degeneracies. To understand the 
difference, recall that the 
frequencies are defined through (\ref{eq:DefOfFrequencies}),
so that each of them is a function of the action variables 
$J_1,\cdots,J_f$. A proper degeneracy holds identical for all  
considered values $J_1,\cdots,J_f$ (which must at least contain 
for each $J_k$ an open interval of values around the considered 
value), whereas an improper degeneracy only holds for singular 
values of the $J$s. This distinction should then also be made 
for the notion of $r$-fold degeneracy: a proper $r$-fold 
degeneracy of frequencies is such that it holds identical for 
a whole neighbourhood of values $J_1,\cdots,J_f$ around the 
considered one.

The possibility of degeneracies and their relevance for the 
formulation of quantisation conditions was already anticipated 
by Schwarzschild~\cite{Schwarzschild:1916c}, who was 
of course very well acquainted with the more refined aspects of 
Hamilton-Jacobi theory, e.g. through Charlier's widely read 
comprehensive treatise 
\cite{Charlier:Himmelsmechanik-1,Charlier:Himmelsmechanik-2}.
Schwarzschild stated in \S3 of \cite{Schwarzschild:1916c} that 
if action-angle variables could be found for which some 
of the frequencies $\nu_k$, say $\nu_{s+1},\cdots,\nu_{s+r}$
where $s+r=f$ vanished, then no quantum condition should be 
imposed on the corresponding actions $J_{s+1},\cdots, J_{s+r}$. 
The rational for that description he gave was that defining 
equation (\ref{eq:DefOfFrequencies}) for the frequencies showed 
that the energy $W$ was independent of $J_1,\cdots,J_k$. 
In his words (and our notation):
\begin{quote}
``This amendment to the prescription [of quantisation] is suggested
by the remark, that for a vanishing mean motion $\nu_k$, the equation 
$\nu_k=\partial W/\partial J_k$ shows that the energy becomes 
independent of the variables $J_k$, that therefore these variables
have no relation to the energetic process within the system.''
(\cite{Schwarzschild:1916c}, p.\,550)
\end{quote}
From that it is clear that the independence of the energy $W$ of 
the $J_k$ for which $\nu_k=0$ is only given if the system is 
\emph{properly} degenerate; otherwise we just have a stationary 
point of $W$ with respect to $J_k$ at that particular $J_k$-value. 
So Schwarzschild's energy argument only justifies to not quantise 
those action variables whose conjugate angles have frequencies 
that vanish identically in the $J_k$ (for some open neighbourhood).

Now, it is true that for a $r$-fold degenerate system (proper 
or improper) a canonical transformation exists so that, say, 
the first $s=f-r$ frequencies $\nu_1,\cdots,\nu_s$ are 
non-degenerate, whereas the remaining $r$ 
frequencies $\nu_{s+1},\cdots,\nu_{s+r}$ are all zero (for the 
particular values of $J$s in the improper case). The number $s$ 
of independent frequencies is called the \emph{degree of periodicity} 
of the system (\cite{Born:VorlesungenAtommechanik}, p.\,105). 
Hence Scharzschild's energy argument amounts to the statement, that 
for \emph{proper degeneracies} only the $s$ action variables 
$J_1,\cdots, J_s$ should be quantised, but not the remaining 
$J_{s+1},\cdots,J_{s+r}$. If the degeneracies are improper, 
systems for arbitrarily close values of the $J_k$ would have them 
quantised, so that it would seem physically unreasonable to treat 
the singular case differently, as Epstein argued in \cite{Epstein:1916c}
in reaction to Schwarzschild. 

Born now proceeds to generalise the uniqueness theorem to degenerate 
systems. For this one needs to determine the most general 
transformations preserving conditions A-C and, in addition, 
the separation into $s$ independent and $r$ mutually 
dependent (vanishing) frequencies. This can indeed be done, 
so that the above theorem has the following natural 
generalisation:     

\medskip
\noindent
\textbf{Theorem (Uniqueness for degenerate systems)}
\emph{
If for a mechanical system variables $(w,J)$ can be introduced 
satisfying conditions A-C, then they can always be chosen in such 
a way that the first $s$ of the partial derivatives 
\begin{equation}
\label{eq:DefOfFrequencies-2}
\nu_k=\frac{\partial W}{\partial J_k}\,,
\end{equation} 
i.e. the $\nu_1,\cdots,\nu_s$ are incommensurable and the others 
$\nu_{s+1},\cdots,\nu_{s+r}$, where $s+r=f$, vanish. Then the 
first $s$ action variables, $J_1,\cdots,J_s$, are determined 
uniquely up to transformations of type 
(\ref{eq:IntegerTransformationActionVar})
[that is, linear transformations by $\mathrm{GL}(s,\integers)$]}.

In the next section (\S\,16), Born completes these results 
by showing that adiabatic invariance holds for $J_1,\cdots,J_s$
but not for $J_k$ for $k>s$, even if the degeneracy is merely 
improper (\cite{Born:VorlesungenAtommechanik} p.\,111). 
He therefore arrives at the following 

\medskip
\noindent
\textbf{Quantisation rule:}
\emph{Let the variables $(w,J)$ for a mechanical system satisfying 
conditions A-C  be so chosen that $\nu_1,\cdots,\nu_s$ are 
incommensurable and $\nu_{s+1},\cdots,\nu_{s+r}$ ($s+r=f$) vanish 
(possibly $r=0$). The stationary motions of this systems are then 
determined by 
\begin{equation}
\label{eq:BornQuantumCond}
J_k=n_kh\qquad\text{for}\quad k=1\,\cdots,s\,.
\end{equation}}
Born acknowledges that Schwarzschild already proposed to 
exempt those action variables from quantisation whose conjugate 
angles have degenerate frequencies. But, at this point, he does 
not sufficiently clearly distinguish between proper and improper 
degeneracies. This issue is taken up again later in Chapter\,4
on perturbation theory, where he states that the (unperturbed) 
system, should it have improper degeneracies, should be quantised 
in the corresponding action variables 
(cf. p.\,303 of \cite{Born:VorlesungenAtommechanik}).

\subsubsection*{A simple system with (proper) degeneracies}
\label{sec:ExamplesDegeneracies}
To illustrate the occurrence of degeneracies, we present in a 
slightly abbreviated form the example of the 3-dimensional 
harmonic oscillator that Born discusses in \S\,14 for the 
very same purpose. Its Hamiltonian reads 
\begin{equation}
\label{eq:HarmOszHamiltonian}
H=
\frac{1}{2m}\bigl(p_1^2+p_2^2+p_3^2\bigr)+
\frac{m}{2}\bigl(\omega_1^2x_1^2+\omega_2^2x_2^2+\omega_3^2x_3^2\bigr)\,.
\end{equation}
The general solution to the Hamilton-Jacobi equation is 
($i=1,2,3$):
\begin{subequations}
\label{eq:HarmMotion}
\begin{alignat}{2}
\label{eq:HarmMotion-a}
&x_i&&\,=\,\sqrt{\frac{J_i}{2\pi^2\nu_i^2m}}\ \sin(2\pi w_i)\,,\\
\label{eq:HarmMotion-b}
&p_i&&\,=\,\sqrt{2\nu_imJ_i\,}\ \cos(2\pi w_i)\,,
\end{alignat}
where
\begin{equation}
\label{eq:HarmMotion-c}
\nu_i=\frac{\omega_i}{2\pi}\quad\text{and}\quad
w_i=\nu_i t+\delta_i\,.
\end{equation}
\end{subequations}
The $\delta_i$ and $J_i$ are six integration constants, in terms
of which the total energy reads 
\begin{equation}
\label{eq:HarmMotionEnergy}
W=\sum_{_i=1}^3 \nu_iJ_i\,.
\end{equation}

Now, a one-fold degeneracy occurs if the frequencies $\nu_i$ obey 
a single relation of the form 
\begin{equation}
\label{eq:HarmDegCond1}
\sum_{_i=1}^3 \tau_i\nu_i=0\,,
\end{equation}
where $\tau_i\in\integers$. This happens, for example, if 
\begin{equation}
\label{eq:HarmDegCond2}
\omega_1=\omega_2=:\omega\ne\omega_3\,, 
\end{equation}
in which case the Hamiltonian is invariant under rotations around 
the third axis. The energy then only depends on $J_3$ and the sum 
$(J_1+J_2)$. Introducing coordinates $x'_i$ with respect to 
a system of axes that are rotated by an angle $\alpha$ around 
the third axis, 
\begin{subequations}
\label{eq:HarmCoordRotation}
\begin{alignat}{2}
\label{eq:HarmCoordRotation-a}
&x'_1&&\,=\,x_1\cos\alpha-x_2\sin\alpha\,,\\
\label{eq:HarmCoordRotation-b}
&x'_2&&\,=\,x_1\sin\alpha+x_2\cos\alpha\,,\\
\label{eq:HarmCoordRotation-c}
&x'_3&&\,=\,x_3\,,
\end{alignat}
\end{subequations}
under which transformation the momenta transform just like the 
coordinates\footnote{\label{foot:TransProp}
Generally, the momenta, being elements of
the vector space dual to the velocities, transform via the 
inverse-transposed of the Jacobian (differential) for the 
coordinate transformation. But for linear transformations 
the Jacobian is just the transformation matrix and orthogonality 
implies that its inverse equals its transpose.}. 
The new action variables, $J'_i$, 
are given in terms of the old $(w_i,J_i)$ by: 
\begin{subequations}
\label{eq:HarmTransActionVar}
\begin{alignat}{2}
\label{eq:HarmTransActionVar-a}
&J'_1&&\,=\,
J_1\cos^2\alpha+J_2\sin^2\alpha
-2\sqrt{J_1J_2}\ \cos(w_1-w_2)\sin\alpha\cos\alpha\,,\\
\label{eq:HarmTransActionVar-b}
&J'_2&&\,=\,
J_1\sin^2\alpha+J_2\cos^2\alpha
+2\sqrt{J_1J_2}\ \cos(w_1-w_2)\sin\alpha\cos\alpha\,,\\
\label{eq:HarmTransActionVar-c}
&J'_3&&\,=\,J_3\,.
\end{alignat}
\end{subequations}
As Born stresses, the $J'_i$ do not just depend on the $J_i$s, but also 
on the $w_i$s, more precisely on the difference $w_1-w_2$, which 
is a constant ($\delta_1-\delta_2$) along the dynamical trajectory 
according to~(\ref{eq:HarmMotion-c}) and (\ref{eq:HarmDegCond2}), 
as it must be (since the $J'_i$ are constant). It is now clear that, 
for general $\alpha$, the conditions $J_{1,2}=n_{1,2} h$ and 
$J'_{1,2}=n'_{1,2} h$ are mutually incompatible. 
However, (\ref{eq:HarmTransActionVar}) shows that the sums 
are invariant  
\begin{equation}
\label{eq:HarmInvSums}
J'_1+J'_2=J_1+J_2\,,
\end{equation}
hence a condition for the sum 
\begin{subequations}
\label{eq:HarmQuantCond1}
\begin{equation}
\label{eq:HarmQuantCond1-a}
J'_1+J'_2=J_1+J_2=n h
\end{equation}
together with
\begin{equation}
\label{eq:HarmQuantCond1-b}
J'_3=J_3=n_3 h
\end{equation}
\end{subequations}
makes sense. 

But what about other coordinate changes than just rotations?
To see what happens, Born considers instead of 
(\ref{eq:HarmCoordRotation}) the transformation to cylindrical 
polar coordinates $(r,\varphi,z)$ with conjugate momenta 
$(p_r,p_\varphi,p_z)$ (cf.~footnote\,\ref{foot:TransProp}):    
\begin{subequations}
\label{eq:HarmCylPolarCoord}
\begin{alignat}{4}
\label{eq:HarmCylPolarCoord-a}
&x_1&&\,=\,r\,\cos\varphi\qquad 
&&p_r&&\,=\,p_1\,\cos\varphi+p_2\,\sin\varphi\,,\\
\label{eq:HarmCylPolarCoord-b}
&x_2&&\,=\,r\,\sin\varphi\qquad 
&&p_\varphi&&\,=-\,p_1r\,\sin\varphi+p_2r\,\cos\varphi\,,\\
\label{eq:HarmCylPolarCoord-c}
&x_3&&\,=\,z\qquad 
&&p_z&&\,=\,p_3\,.
\end{alignat}
\end{subequations}
The transformation equations from the old $(w_i,J_i)$ to the new 
action variables $(J_r,J_\varphi,J_z)$ are%
\footnote{In Born's text \cite{Born:VorlesungenAtommechanik}, p.\,98, 
there are additional factors $\nu^{-1}$ in front of $\sqrt{J_1J_2}$ 
in \eqref{eq:HarmTransActionVar2-a} as well as \eqref{eq:HarmTransActionVar2-b}. These are incorrect,
as can be told already on dimensional grounds.}:
\begin{subequations}
\label{eq:HarmTransActionVar2}
\begin{alignat}{2}
\label{eq:HarmTransActionVar2-a}
&J_r&&\,=\,\tfrac{1}{2}(J_1+J_2)-
        \sqrt{J_1J_2}\,\sin\bigl(2\pi(w_1-w_2)\bigr)\,,\\
\label{eq:HarmTransActionVar2-b}
&J_\varphi&&\,=\,2\sqrt{J_1J_2}\,\sin\bigl(2\pi(w_1-w_2)\bigr)\,,\\
\label{eq:HarmTransActionVar2-c}
&J_z&&\,=\,J_3\,.
\end{alignat}
\end{subequations}
The total energy expressed as a function of the new action 
variables reads: 
\begin{equation}
\label{eq:HarmMotionEnergy2}
W=\nu(2J_r+J_\varphi)+\nu_zJ_z\,,
\end{equation}
where here and in (\ref{eq:HarmTransActionVar2})
$\nu:=\omega/2\pi$ and $\nu_z:=\omega_3/2\pi$ (cf. (\ref{eq:HarmDegCond2})). 
Again it is only the combination $2J_r+J_\varphi$ that enters the energy 
expression, and from (\ref{eq:HarmTransActionVar2}) we see immediately 
that that
\begin{equation}
  \label{eq:HarmActionVarIdentity}
2J_r+J_\varphi=J_1+J_2\,,
\end{equation}
Again, conditions of the form 
$J_r=n_r h$, $J_\varphi=n_\varphi h$, and  $J_z=n_z h$ 
would pick out different  ``quantum orbits'' 
[Born speaks of ``Quantenbahnen''] than those corresponding 
to $J_i=n_i h$. The energies, however, are the same.

\section{Einstein's view}
Already in 1917 Einstein took up the problem of 
quantisation in his long neglected\footnote{Einsteins 
paper was cited by de\,Broglie in his thesis 
\cite{Broglie:1925}, where he spends slightly more than 
a page (pages 64-65 of Section\,II in Chapter\,III) to 
discuss the ``interpretation of Einstein's quantisation 
condition'', and also in Schr\"odinger's \emph{Quantisation 
as Eigenvalue Problem}, where in the Second Communication 
he states in a footnote that Einstein's quantisation 
condition ``amongst all older versions stands closest to the 
present one [Schr\"odinger's]''. However, after Matrix- 
and Wave Mechanics settled, Einstein's paper seems to 
have been largely forgotten until Keller~\cite{Keller:1958} 
reminded the community of its existence in 1958.} paper  
``On the Quantum Theorem of Sommerfeld and Epstein'' 
(\cite{Einstein:CP}, Vol.\,6, Doc.\,45, p.\,556-567). 
Einstein summarised this paper in a letter to Ehrenfest
dated June\,3rd 1917 (\cite{Einstein:CP}, Vol.\,8, Part\,A, 
Doc.\,350, pp.\,464-6), in which he also makes very 
interesting comments, as we shall see below. 
For discussions of its content from a modern viewpoint 
see, e.g., \cite{Gutzwiller:ChaosClassQuantPhys} and 
\cite{Stone:2005}.)

In this paper Einstein suggested to replace the quantum condition 
(\ref{eq:QuantRuleManyDegFreed}) by 
\begin{equation}
\label{eq:EinsteinsQuantCond}
\oint_\gamma \sum_{k=1}^f p_k dq_k=n_\gamma h\,,\qquad \forall \gamma\,.
\end{equation}
First of all one should recognise that here the sum rather 
than each individual term $p_k dq_k$ as in 
(\ref{eq:QuantRuleManyDegFreed}) forms the integrand. 
Second, (\ref{eq:EinsteinsQuantCond}) is not just one but 
many conditions, as many as there are independent paths 
(loops) $\gamma$ against which the integrand is integrated.

Let us explain the meaning of all this in a modernised 
terminology. For this, we first point out that the integrand 
has a proper geometric meaning, since   
\begin{equation}
\label{eq:LiouvilleForm}
\theta=\sum_{k=1}^f p_k dq_k
\end{equation}
is the coordinate expression of a global one-form on 
phase space (sometimes called the Liouville form)%
\footnote{In the terminology of differential geometry, phase 
space is the cotangent bundle $T^*Q$ over configuration space $Q$
with projection map $\pi:T^*Q\rightarrow Q$. The one-form 
$\theta$ on $T^*Q$ is defined by the following rule: 
Let $z$ be a point in $T^*Q$ and $X_z$ a vector in the 
tangent space of $T^*Q$ at $z$, then 
$\theta_z(X_z):=z\bigl(\pi_*\vert_z(X_z)\bigr)$. 
Here the symbol on the right denotes the differential of 
the projection map $\pi$, evaluated at $z$ and then applied 
to $X_z$. This results in a tangent vector at $\pi(z)$ on $Q$
on which $z\in T_{\pi(z)}^*Q$ may be evaluated. In local 
adapted coordinates $(q_1,\cdots,q_f;p_1,\cdots,p_f)$ 
the projection map $\pi$ just projects onto the $q$s.
Then, for $X=\sum_k (Y_k\partial_{q_k}+Z_k\partial_{p_k})$
we have $\pi_*(X)=\sum_kY_k\partial_{q_k}$ and 
$z\bigl(\pi_*(X)\bigr)=\sum_kp_kY_k$, so that 
$\theta=\sum_kp_k\,dq_k$.}, quite in 
contrast to each individual term $p_k\,dq_k$, which has no 
coordinate independent geometric meaning. Being a one-form 
it makes invariant sense to integrate it along paths.
The paths $\gamma$ considered here are all closed, i.e. loops, 
hence the $\oint$-sign. But what are the loops $\gamma$ 
that may enter (\ref{eq:EinsteinsQuantCond})? For their 
characterisation it is crucial to assume that the system 
be integrable. This means that there are $f$ ($=$ number 
of degrees of freedom) functions on phase space, $F_A(q,p)$ 
($A=1,\cdots, f$), the energy being one of them, whose 
mutual Poisson brackets vanish: 
\begin{equation}
\label{eq:CommConstants}
\{F_A,F_B\}=0\,.
\end{equation}
This implies that the trajectories remain on the level sets 
for the $f$-component function $\vec F=(F_1,\cdots,F_f)$, 
which can be shown to be $f$-dimensional tori $T_{\vec F}$ 
embedded in  $2f$-dimensional phase space. From 
(\ref{eq:CommConstants}) it follows that these tori are 
geometrically special (Lagrangian) submanifolds: The 
differential of the one form (\ref{eq:EinsteinsQuantCond}),
restricted to the tangent spaces of these tori, vanishes 
identically. By Stokes' theorem this implies that any two 
integrals of $\theta$ over loops $\gamma$ and $\gamma'$ 
within the same torus $T$ coincide in value (possibly up 
to sign, depending on the orientation given to the loops) 
if there is a 2-dimensional surface $\sigma$ within $T$ 
whose boundary is just the union of $\gamma$ and $\gamma'$. 
This defines an equivalence relation on the set of loops 
on $T$ whose equivalence classes are called homology 
classes (of dimension 1). The homology classes form a 
finitely generated Abelian group (since the level sets
are compact) so that each member can be uniquely written 
as a linear combination of $f$ basis loops (i.e. their classes) 
with integer coefficients. For example, if one pictures the 
$f$-torus as an $f$-dimensional cube with pairwise identifications
of opposite faces through translations, an $f$-tuple of basis loops 
is represented by the straight lines-segments connecting the 
midpoints of opposite faces. Each such basis is connected 
to any other by a linear $\mathrm{GL}(f,\integers)$ transformation. 

Now we can understand how (\ref{eq:EinsteinsQuantCond}) 
should be read, namely as a 
condition that selects out of a continuum a discrete 
subset of tori $T_{\vec F}$, which may be characterised 
by discretised values for the $f$ observables 
$F_A$. By the last remark of the previous paragraph it 
does not matter which basis for the homology classes of 
loops is picked to evaluate (\ref{eq:EinsteinsQuantCond}).
This leads to quantisation condition independent of the 
need to separate variables.

What remains undecided at this stage is how to proceed 
in cases where degeneracies occur. In the absence of degeneracies,  
the torus is \emph{uniquely} determined as the closure of 
the phase-space trajectory for all times. If degeneracies 
exist, that closure will define a torus of dimension  $s<f$, 
the embedding of which in a torus of dimension $f$ is ambiguous 
since the latter is not uniquely determined by the motion 
of the system. This we have seen by Born's examples above. 
Even simpler examples would be the planar harmonic oscillator 
and planar Keplerian motion; cf. Sect,\,51 of 
\cite{Arnold:Mechanics}). In that case one has to decide 
whether (\ref{eq:EinsteinsQuantCond}) is meant to apply only 
to the $s$ generating loops of the former or to all $f$ of 
the latter, thus introducing an $f-s$ fold ambiguity in the 
determination of ``quantum orbits'' [Born: ``Quantenbahnen''].

The geometric flavour of these arguments are clearly present in 
Einstein's paper, though he clearly did not use the modern 
vocabulary. Einstein starts from the $f$-dimensional configuration 
space that is coordinatised by the $q$s and regards the 
$p$s as certain `functions' on it, defined through an
$f$ parameter family of solutions. Locally in $q$-space 
(i.e. in a neighbourhood or each point) Hamilton's equations 
guarantee the existence of ordinary (i.e. single valued) 
functions $p_k(q_1,\cdots,q_f)$. However, following a dynamical 
trajectory that is dense in a portion of $q$-space the 
values $p_k$ need not return to their original values. 
Einstein distinguishes between two cases: either the number 
of mutually different $p$-values upon return of the trajectory 
in a small neighbourhood $U$ around a point in $q$-space is 
finite, or it is infinite. In the latter case Einstein's 
quantisation condition does not apply. In the former case, 
Einsteins considers what he in the letter to Ehrenfest called 
the Riemannianisation (``Riemannisierung'') of $q$-space, 
that is, a finite-sheeted covering. The components $p_k$ 
will then be a well defined (single valued) co-vector field 
over the dynamically allowed portion of $q$-space
(see \cite{Stone:2005} for a lucid discussion with pictures). 

In a most interesting 1.5-page supplement added in proof, Einstein 
points out that the first type of motion, where $q$-space 
trajectories return with infinitely many mutually different 
$p$-values, may well occur for simple systems with relatively 
few degrees of freedom, like e.g. that of three pointlike 
masses moving under the influence of their mutual gravitational 
attractions, as was first pointed out by Poincar\'e in the 
1890s to which Einstein refers. Einstein ends his supplement 
(and the paper) by stating that for non-integrable systems 
his condition also fails. In fact, as discussed above, it even
cannot be written down. 

Hence one arrives at the conclusion that the crucial question 
concerning the applicability of quantisation conditions is that 
of \emph{integrability}, i.e. whether sufficiently many constants 
of motion exist; other degrees of complexity, like the number 
of degrees of freedom, do not directly matter. As we know from 
Poincar\'e's work, non-integrability occurs already at the 
3-body level for simple 2-body interactions. But what is 
the meaning of ``Quantum Theory'' if ``quantisation''
is not a universally applicable procedure?\footnote{Even today 
this question has not yet received a unanimously accepted answer.}

%Here are Einstein's own words in the letter to Ehrenfest:
%
%\begin{quote}
%``Consider a mechanical problem for which there exist as many 
%integrals $L(q,p)=const.$ as degrees of freedom. (Here $(q,p)$ 
%means that $L$ depends an all coordinates $q^\nu$ and momenta 
%$p_^\nu$ of phase space.) Then the momenta $p_\nu$ can be 
%expresses as multiply-valued functions of the $q_\nu$. 
% On the other hand, let the trajectory (Bahnkurve) fill up 
% a certain part of $q$-spaces completely, so that it approaches 
% every point in it arbitrarily closely. Then the orbit (Bahn) 
% of the System gives a vector field for the $p$'s in $q$-space. 
% In a $q$-space that is foliated according to Riemann ('Riemannisch 
% gebl\"attert') the $p$ can be interpreted as \emph{unique} functions
%of the $q$. Now consider the sum $d\sigma=\sum_\nu p_\nu\,dq^\nu$ for an 
%arbitrary line-element of $q$-space. It is invariant under coordinate
%transformations and, in addition, a \emph{complete differential}
%(it is closed but not exact 1-form). The latter can be deduced from 
%Jacobi's theorem. 
%
%Now consider the integral $\int d\sigma$, taken along a closed 
%curve in $q$-space that is foliated according to Riemann. 
%It vanishes if the curve can be contracted continuously to a point,
%which may well not be the case for all curves due to Riemannisation
%("Riemannisierung"). Now, the quantum rule simply requires that for 
%\emph{each} closed curve $\int\sum_\nu p_\nu\,dq^\nu=nh$. The special 
%case of Epstein ist just that, where each $p_\nu$ merely depends on 
%the associated $q_\nu$.      
In the letter to Ehrenfest already mentioned above, Einstein 
stresses precisely this point, i.e. that his condition is only 
applicable to integrable systems, and ends with a truly 
astonishing statement (here the emphases are mine):
\begin{quote}
``As pretty as this may appear, it is just restricted to 
the special case where the $p_\nu$ can be represented as 
(multi-valued) functions of the  $q_\nu$. It is interesting 
that this restriction just nullifies the validity of statistical 
mechanics. The latter presupposes that upon recurrence of the 
$q_\nu$, the $p_\nu$ of a system in isolation assume all values 
by and by which are compatible with the energy principle.
\emph{It seems to me, that the true [wirkliche] mechanics is such 
that the existence of the integrals (which exclude the validity 
of statistical mechanics) is already assured by the general 
foundations.} But how to start??''\footnote{%
``So h\"ubsch nun diese Sache ist, so ist sie eben auf den 
Spezialfall beschr\"ankt, dass die $p_\nu$ als (mehrdeutige) 
Funktion der $q_\nu$ dargestellt werden k\"onnen. Es ist 
interesstant, dass diese Beschr\"ankung gerade die G\"ultigkeit 
der statistischen Mechanik aufhebt. Denn diese setzt voraus, 
dass die $p_\nu$ eines sich selbst \"uberlassenen Systems bei 
Wiederkehr der $q_\nu$ nach und nach alle mit dem Energieprinzip 
vereinbaren Wertsysteme annehmen. \emph{Es scheint mir, dass die 
wirkliche Mechanik so ist, dass die Existenz der Integrale, 
(welche die G\"ultigkeit der statistischen Mechanik ausschliessen), 
schon verm\"oge der allgemeinen Grundlagen gesichert ist.} 
Aber wie ansetzen??."}
(\cite{Einstein:CP}, Vol.\,8, Part\,A, Doc.\, 350, p.\,465)
\end{quote}
Are we just told that Einstein contemplated the impossibility 
of any rigorous foundation of classical statistical 
mechanics?

\section{Final comments}
In his book, Born also mentions Poincar\'e's work and cites 
the relevant chapters  on convergence of perturbation series 
and the 3-body problem in Charlier's 
treatise~\cite{Charlier:Himmelsmechanik-2}, but he does not 
seem to make the fundamental distinction between integrable 
and non-integrable systems in the sense Einstein made it.
Born never cites Einstein's paper in his book. 
He mentions the well known problem (since Bruns 1884) of 
small denominators (described in Chapter\,10, \S\,5 of 
\cite{Charlier:Himmelsmechanik-2}) and also 
Poincar\'e's result on the impossibility to describe the 
motion for even arbitrarily small perturbation functions in
terms of convergent Fourier series. From that Born concludes 
the impossibility to introduce constant $J_k$s and hence the 
impossibility to pose quantisation rules in general.
His conclusion from that is that, for the time being, 
one should take a pragmatic attitude (his emphases):
\begin{quote}
``Even though the mentioned approximation scheme does not 
converge in the strict sense, it has proved useful in 
celestial mechanics. For it could be shown [by Poincar\'e] 
that the series showed a \emph{type of semi-convergence}.
If appropriately terminated they represent the motion of
the perturbed system with great accuracy, not for 
arbitrarily long times, but still for practically very 
long times. \emph{From this one sees on purely theoretical 
grounds, that the absolute stability of atoms cannot be 
accounted for in this way}. However, for the time being 
one will push aside [sich hinwegsetzen] this fundamental 
difficulty and make energy calculations test-wise, in order 
to see whether one obtains similar agreements as in celestial 
mechanics.'' (\cite{Born:VorlesungenAtommechanik}, p.\,292-293)
\end{quote}
    
Ten pages before that passage, in the introduction to the chapter 
on perturbation theory, Born stressed the somewhat ambivalent
situation perturbation theory in atomic physics faces in
comparison to celestial mechanics: One one hand, `perturbations' 
caused by electron-electron interactions are of the same order of
magnitude than electron-nucleus interactions, quite in contrast 
to the solar system , where the sun is orders of magnitude 
heavier than the planets. On the other hand, the quantum 
conditions drastically constrain possible motions and could well 
act as regulator. As regards the analytical difficulties 
already mentioned above, he comments in anticipation:
\begin{quote}
``Here [convergence of Fourier series] an insurmountable
analytical difficulty seems to inhibit progress, and one 
could arrive at the opinion that it is impossible to gain 
a theoretical understanding of atomic structures up to 
Uranium.''
(\cite{Born:VorlesungenAtommechanik}, p.\,282-283)
\end{quote}
However,
\begin{quote}
``The aim of the investigations of this chapter shall be to 
demonstrate, that this is difficulty is not essential. It 
would indeed be strange [sonderbar] if Nature barricaded
herself behind the analytical difficulties of the $n$-body 
problem against the advancement of knowledge [das Vordringen
der Erkenntnis].''
(\cite{Born:VorlesungenAtommechanik}, p.\,282-283)
\end{quote}

In the course of the development of his chapter on perturbation 
theory very interesting technical points come up, one of them 
being connected with the apparent necessity to impose 
quantisation conditions for the unperturbed action variables 
conjugate to angles whose frequencies are \emph{improperly} 
degenerate. But the discussion of this is quite technical and 
extraneous to Born's approach to the quantisation procedure.

%\addcontentsline{toc}{section}{References}
\bibliographystyle{plain}
\bibliography{RELATIVITY,HIST-PHIL-SCI,MATH,QM} 
\end{document}